\documentclass[12pt]{article}
\usepackage{amssymb,amsmath,amscd,amsthm}
\usepackage{graphicx,psfrag,epsfig, color,float}

\usepackage{graphicx}
\usepackage[active]{srcltx}

\newtheorem{theorem}{Theorem}[section]

\newtheorem{lemma}[theorem]{Lemma}

\date{}

\begin{document}

\title{The Radius of a Polymer at a Near-Critical Temperature}
\author{ L.
Koralov\footnote{Dept of Mathematics, University of Maryland,
College Park, MD 20742, koralov@math.umd.edu}, S.
Molchanov\footnote{Dept of Mathematics and Statistics, UNCC, NC 28223 and National Research University, Higher School of Economics, Russian Federation,
smolchan@uncc.edu}, B. Vainberg\footnote{Dept of Mathematics and Statistics, UNCC, NC 28223, brvainbe@uncc.edu} }
\maketitle

\vskip1cm
\begin{abstract}
We consider a mean-field model of a polymer with a spherically-symmetric finitely
supported potential.
We
describe how the typical size of the polymer depends on the two
parameters: the temperature, which
approaches the critical value, and the length of the polymer chain, which goes to infinity.
\end{abstract}


\section{Introduction}

In a simple model, a polymer can be considered as a sequence of monomers (small molecules) connected into a chain, with
consecutive edges forming random angles with each other. We assume that the origin is a starting point of the polymer chain. When the number of monomers is large (and their sizes are small), the chain can be viewed as
a realization of a continuous time random process on the interval $[0,t]$, where $t$ is proportional
to the number of the monomers.  The distribution
of the process (i.e., the measure on the space of realizations) depends on the temperature (in fact, we will use the inverse value of the temperature, which
will be denoted by $\beta$). At infinite temperature, i.e., for $\beta =0$, the direction of each edge in the chain of monomers is independent of
the rest, resulting in a Brownian motion as the limiting (continuous-time) model for the realizations of the polymer.

For positive values of $\beta$, one needs to take  the interaction between the monomers into account. In this paper, we will use
the mean field-type model (also called the deterministic pinned model), introduced by Lifshitz, Grosberg, and Khokhlov (\cite{LGK}), in which the
interactions between the individual monomers are replaced by their interaction with an attracting external potential.
More precisely, the distribution in the continuous-time polymer model is given by the Gibbs measure  $\mathrm{P}_{\beta,t}$ with a nonnegative, not
identically equal to zero potential  $v \in C_0^{\infty}(\mathbb{R}^d)$  and the  coupling constant $\beta \geq 0$ equal to the inverse temperature (regulating the strength of the attraction), that is
\[
\frac{d\mathrm{P}_{\beta,t}}{d\mathrm{P}_{0,t}}(\omega)=
\frac{e^{\beta\int_0^tv(\omega(s))ds}}{Z_{\beta,t}},\qquad
\omega\in C([0,t],\mathbb{R}^d),
\]
where $\mathrm{P}_{0,t}$ is the Wiener measure on the space $C([0,t],\mathbb{R}^d)$ and
\[
Z_{\beta,t}=\mathrm{E}_{0,t}e^{\beta\int_0^tv(\omega(s))ds}
\]
is the partition function.
In other words, one starts with the Wiener measure $\mathrm{P}_{0,t}$,
constructs a measure with the density $e^{\beta\int_0^tv(\omega(s))ds}$ with respect to $\mathrm{P}_{0,t}$, and normalizes it to make it a probability measure.

At the critical value of the temperature, polymers exhibit a transition between the folded (globular) and unfolded states.
Namely, under the measure $\mathrm{P}_{0,t}$, $\omega(\cdot)$ is simply a trajectory of a Brownian motion, and thus $|\omega(t)|$ is typically of order $\sqrt{t}$.
The behavior is qualitatively similar under the measure $\mathrm{P}_{\beta,t}$ for small $\beta$.
 When
$\beta$ is sufficiently large (larger than a certain value $\beta_{\rm cr}$), under the measure $\mathrm{P}_{\beta,t}$,
$\omega(\cdot)$ converges, as $t \rightarrow \infty$, to a Markov process that
has an invariant probability distribution, and thus $|\omega(t)|$ is typically of order one.
 We define the radius of the polymer
as
\begin{equation} \label{dra}
r(\beta, t) = (\mathrm{E}_{\beta, t} |\omega(t)|^2)^{\frac{1}{2}}.
\end{equation}
The value $\beta_{\rm cr}$ separates the two types of the limiting behavior of the radius. Namely, $r(\beta, t) \sim c_\beta \sqrt{t}$ as $t \rightarrow \infty$
for $\beta < \beta_{\rm cr}$, while $r(\beta, t) \rightarrow \tilde{c}_\beta$ as $t \rightarrow \infty$
for $\beta > \beta_{\rm cr}$.
The situation is most interesting when $\beta$ approaches $\beta_{\rm cr}$ and
$t$ goes to infinity, simultaneously. We will examine how the size of the polymer depends on these two parameters in this situation.
Let us stress that, in applications, $t$ is large but far from infinite. Unlike other types of phase transitions (e.g., in magnetization, where the
number of spins can be of order $10^{24}$), the number of monomers comprising a protein is usually of order $10^4$ or less. At the same time, the range of values
of $\beta$ where the transition occurs is very narrow, and it is difficult to experimentally measure the dependence of the quantities
such as $r(\beta, t)$ on $\beta$ near the critical value. This explains the importance of the two-parameter asymptotics.
\\
{\bf Remark.} The random quantity $\left(\frac{1}{t} \int_0^t |\omega(s)|^2 ds\right)^{\frac{1}{2}}$
could be taken as an alternative definition of the radius. Due to the averaging, this quantity behaves nearly deterministically for large $t$, namely, as
$\left(\frac{1}{t} \int_0^t  \mathrm{E}_{\beta, t}|\omega(s)|^2 ds\right)^{\frac{1}{2}}$, which, in turn,
is asymptotically equivalent (up to a positive constant factor) to $r(\beta, t) $.

In our model we don't take the time fluctuations of the polymer chain into account since the time fluctuations of the effective radius are assumed to be negligible
due to the large length of the polymer. Thus our model is time-independent. However, when studying the distribution of the polymer, the distance along the polymer will play the same role as the time variable in parabolic equations, hence the notation~$t$.
\\


From the Feynman-Kac formula it follows that, under the
measures $\mathrm{P}_{\beta,t}$, the finite-dimensional distributions of the process $\{\omega(s), 0\leq s\leq
t\}$ can be expressed in terms of the fundamental
solution $p_{\beta}$ of the parabolic equation
\begin{equation} \label{frst}
\frac{\partial u}{\partial t}=H_{\beta}u, ~~~{\rm where}~~~~
H_{\beta}=\frac{1}{2}\Delta  +\beta v: L^2(\mathbb{R}^d)\to
L^2(\mathbb{R}^d).
\end{equation}
In particular,
\begin{equation} \label{smom}
r^2(\beta, t) = \frac{\int_{\mathbb{R}^3} |x|^2 p_{\beta}(t,0,x)dx }{  \int_{\mathbb{R}^3}p_{\beta}(t,0,x)dx},
\end{equation}
where $p_{\beta}(t,0,x)$ is the solution of problem (\ref{frst}) with initial data $p_{\beta}|_{t=0}=\delta_0(x)$.

It was shown in \cite{CKMV09} that for $d\geq 3$ the critical value $\beta_{\rm cr}$ of the coupling
constant
coincides  with the spectral bifurcation point for operator $H_\beta$: the spectrum is absolutely continuous and coincides with semi-axis $(-\infty,0]$ when $\beta<\beta_{\rm cr}$, and additional positive eigenvalues exist when $\beta>\beta_{\rm cr}$, see also \cite{PV}. Thus
\[
\beta_{\rm cr}=\sup\{\beta>0|\sup\sigma(H_{\beta})=0\},
\]
where  $\sigma(H_{\beta})$ is the spectrum of the operator  $ H_{\beta}$.

We'll focus on the case when $d = 3$. In order to make the calculations more explicit, we'll assume a particular form of the potential (a similar result
for general potentials will require the asymptotic analysis of the resolvent of $H_\beta$ (see \cite{V75}) and will be published elsewhere). Namely, we'll assume that
\[
v(x) =  \left\{\begin{array}{c}
                            1~~~ {\rm if} ~~ | x | \leq 1,\\
                            0~~~{\rm if} ~~ |x| > 1.
                          \end{array}
													\right.
\]

For two functions $f$ and $g$, we'll write $f \approx g$ if
there are positive constants $c$ and $C$ such that $cf \leq g \leq Cf$ for all the values of the variables that are sufficiently close
to their asymptotic values (in our case, $\beta$ is sufficiently close to $\beta_{\rm cr}$  and $t$ is sufficiently large).

The main result of the paper is the following.
\begin{theorem} \label{mtee}
The radius of the polymer has the following asymptotic
behavior when $|\beta - \beta_{\rm cr}|$ is small and $t$ is large:
\begin{equation} \label{th}
r(\beta, t) \approx \left\{\begin{array}{c}
                            {(\beta-\beta_{\rm cr})^{-1}} ~~~ if ~~ (\beta-\beta_{\rm cr})\sqrt {t}\geq1,\\
                            ~~~~\sqrt{t}~~~~~~~~~if ~~ (\beta-\beta_{\rm cr})\sqrt {t}\leq1 .
                          \end{array}
													\right.
\end{equation}
\end{theorem}
\noindent
{\bf Remark.} The arguments in the proof allow one to specify the coefficients in the asymptotic formula above for every specific
relationship between $\beta - \beta_{\rm cr}$ and $t$.
\\

In \cite{CKMV09}, we used the detailed analysis of the spectral structure of partial differential operators with a compactly supported potential to
describe the distribution of long polymer chains for each fixed value of $\beta$, including $\beta_{\rm cr}$. Subsequently, our results were
generalized and adapted to several related models: the case of power-law decay of the potential at infinity (Lacoin \cite{La}), the case of the
underlying operator being the generator of a stable process (Takeda, Wada \cite{TW}, Li, Li \cite{LiLi}, Nishimori \cite{Ni}), the case of zero-range potentials (our own work \cite{CKMV10}, \cite{KP}, Fitzsimmons, Li \cite{FL}), etc.

Let us contrast the result of the current paper with the results of the closely related work \cite{KP}.
In \cite{KP}, we considered the situation when $\beta = \beta(t)$ is such that
\begin{equation} \label{rela}
(\beta(t)-\beta_{\rm cr})\sqrt{t} \rightarrow \chi \in \mathbb{R}~~~{\rm as}~~{t \rightarrow \infty}.
\end{equation}
 It was shown
that, after scaling the time by $t$ and the spatial variables by $\sqrt{t}$,
the measures $\mathrm{P}_{\beta(t),t}$ converge, as $t \rightarrow \infty$, to certain limiting measures.
The limiting measures~$ \mathrm{Q}_{\chi}$ were introduced in
\cite{CKMV10} as the polymer measures  on $C([0,1],\mathbb{R}^3)$
corresponding to zero-range attracting potentials (i.e., the
potentials that are, roughly speaking, concentrated at the
origin). In the current paper, we do not make the assumption (\ref{rela}), and the two parameters $\beta$ and $t$ can vary independently.
The scaling required to get a nontrivial limit now depends on the relationship between the parameters.

Let us note an important physical implication of our result. When the polymer molecules are observed in a liquid with sufficiently low Reynolds number, the radius affects the diffusion coefficient of the
molecule (denoted by $D(\beta,t)$). Namely, if the molecule in its folded or nearly-folded state is modeled by a ball,  the relationship
between $D(\beta,t)$ and $r(\beta,t)$ is provided by the Stokes-Einstein equation,
\[
D(\beta,t) = \frac{c}{\beta r(\beta,t)}~,
\]
where $c$ is a constant determined by the properties of the media.  This formula allows one to predict the behavior of polymers in a liquid and also provides an approach to study $r(\beta, t)$ experimentally, by measuring
$D(\beta,t)$
(\cite{WGS}, \cite{MFN}).

\section{Proof of the main result}

Solving (\ref{frst}) for $p_\beta$ (with initial data $p_{\beta}(0,0,x)=\delta_0(x)$) using the Laplace transform in  $t$, we get
\begin{equation} \label{lap}
p_\beta(t,0,x)=\frac{1}{2\pi i}\int_{a-i\infty}^{a+i\infty}e^{t\lambda}u(\lambda,r)d\lambda, \quad r=|x|,
\end{equation}
where $a$ is a positive constant and $u\in L^2(\mathbb{R}^3)$ is a spherically symmetric solution of the equation
\begin{equation} \label{uua}
(\frac{1}{2}\Delta  +\beta v-
\lambda)u=-\delta_0(x).
\end{equation}
One can choose any constant $a$ in (\ref{lap}) that is larger than the supremum of the spectrum $\sigma(H_\beta)$ of operator $H_\beta$. It was shown in \cite{CKMV09} that $\sigma(H_\beta)=(-\infty,0]$ when $\beta<\beta_{{\rm cr}}$ and $\sigma(H_\beta)=(-\infty,0]\bigcup \lambda_0$ when $\beta-\beta_{{\rm cr}}>0$ is small enough, where $\lambda_0=\lambda_0(\beta)$ is a simple positive eigenvalue of $H_\beta$ and $\lambda_0(\beta)\to 0$ as $\beta\downarrow\beta_{{\rm cr}}$. Moreover,
there is $\varkappa > 0$ such that
\begin{equation} \label{l0a}
\lambda_0(\beta)\sim \varkappa(\beta-\beta_{{\rm cr}})^2 ~~~{\rm as} ~~~ \beta\downarrow\beta_{{\rm cr}}.
\end{equation}
Thus one can choose, for example, $a=1$ if $|\beta-\beta_{{\rm cr}}|$ is small enough.

Under our assumptions on $v$, the solution $u$ of (\ref{uua}) has the form
\begin{equation} \label{stat}
u=\left\{\begin{array}{c}
                            \frac{\cos{\sqrt{2(\beta-\lambda)}r}}{2\pi r}+B\frac{\sin\sqrt{2(\beta-\lambda)}r}{ \sqrt{2(\beta-\lambda)}r} ~~{\rm when} ~~ r<1\\
                            ~~~~A\frac{e^{-\sqrt{2\lambda} r}}{r}~~~~{\rm when} ~~ r\geq1.
                          \end{array}
													\right.
\end{equation}
Here and below, we assume that $\beta>0$ and that the value of $\sqrt{2 \lambda}$ is chosen so that ${\rm Re}(\sqrt{ 2 \lambda}) > 0$ ($\lambda$ does not
belong to the negative semiaxis). The function $u$ does not have a branching point at $\lambda = \beta$ and its value at $\lambda = \beta$ is defined by continuity.
 The continuity conditions for $u$ on the sphere $r=1$ imply
\begin{equation} \label{sys1}
Ae^{-\sqrt{2\lambda}}-B\frac{\sin\sqrt{2(\beta-\lambda)}}{\sqrt{2(\beta-\lambda)}}-\frac{1}{2\pi}\cos{\sqrt{2(\beta-\lambda)}}=0,
\end{equation}
\begin{equation} \label{sys2}
-A\sqrt{2\lambda}e^{-\sqrt{2\lambda}}-B\cos\sqrt{2(\beta-\lambda)}
+\frac{\sqrt{2(\beta-\lambda)}}{2\pi}\sin{\sqrt{2(\beta-\lambda)}}=0.
\end{equation}
 We will write the determinant $D$ of the system above in the form $D=-e^{-\sqrt{2\lambda}}d$, where
 \begin{equation} \label{dd}
d=d(\beta, \sqrt{\lambda}):=\frac{\sqrt{\lambda}}{\sqrt{\beta-\lambda}}\sin\sqrt{2(\beta-\lambda)}+\cos\sqrt{2(\beta-\lambda)}.
\end{equation}

Solving (\ref{sys1}), (\ref{sys2}), we obtain the following formula for $u$:
\begin{equation} \label{uu}
u=\frac{
\cos[\sqrt{2(\beta-\lambda)}(r-1)]-\frac{\sqrt{\lambda}}{\sqrt{\beta-\lambda}}\sin[\sqrt{2(\beta-\lambda)}(r-1)]}{2\pi rd(\beta, \lambda)}, \quad  r\leq1,
\end{equation}
\begin{equation} \label{uu2}
u=\frac{
e^{-\sqrt{2\lambda}(r-1)}}{2\pi rd(\beta, \lambda)}, \quad  r\geq1.
\end{equation}

Observe that the eigenvalues of the operator $H_\beta$ coincide with zeroes of function $d$.
This is seen by considering the solution of equation (\ref{uua}) with zero in the right hand side in the form (\ref{stat}) without the
term with the cosine function.
Hence, the following statement (proved below) could be equivalently formulated in terms of the eigenvalues of $H_\beta$ instead of the zeros of $d$.
\begin{lemma} The following statements are valid:

1) If $\beta$ is real, then $d\neq0$ for complex (non-real) $\lambda$.

2) If $\beta<\frac{\pi^2}{8}$, then $d\neq0$ when $\lambda\geq0$.

3) There is an $\varepsilon>0$ such that equation $d=0$ has a unique non-negative root $\lambda=\lambda_0(\beta)$ when $\frac{\pi^2}{8}\leq\beta\leq\frac{\pi^2}{8}+\varepsilon$. Function $\lambda=\lambda_0(\beta)$ is analytic on the interval $(\frac{\pi^2}{8}, \frac{\pi^2}{8}+\varepsilon)$, and $\lambda_0(\beta)\sim \frac{1}{2}(\beta-\frac{\pi^2}{8})^2$ as $\beta\downarrow\frac{\pi^2}{8}$.

4) Function $\gamma=\sqrt{\lambda_0(\beta)}$ admits an analytic continuation in a neighborhood of $\beta=\frac{\pi^2}{8}$. 
\end{lemma}
\noindent
{\bf Remark.} 1) In particular, this lemma implies that, under our assumptions on $v$, we have: $\beta_{{\rm cr}}=\frac{\pi^2}{8}$
and constant $\varkappa$ in (\ref{l0a}) equals $1/2$.


2) We may assume (reducing $\varepsilon$ if needed) that
\begin{equation} \label{lest}
\frac{1}{2}|\beta- \beta_{\rm cr}|<|\sqrt{\lambda_0(\beta)}|<|\beta- \beta_{\rm cr}| \quad {\rm when }~~~|\beta- \beta_{\rm cr}|\leq \varepsilon.
\end{equation}

\proof
We assume below that $\beta$ is real. The first statement is obvious since the operator $H_\beta$ is symmetric.

Since the operator $\Delta$ is non-positive and $v(x)\leq 1$, the operator $H_\beta=\frac{1}{2}\Delta+\beta v$ can not have eigenvalues larger than $\beta$. The eigenvalue can  not also be equal to $\beta$ since otherwise the relation $<H_\beta \psi,\psi>=\lambda\|\psi\|^2$ implies that the support of the eigenfunction $\psi$ belongs to the sphere $|x|\leq1$ and that the eigenfunction satisfies the equation $\frac{1}{2}\Delta \psi=0$, which does not have non-trivial compactly supported solutions. Thus $\lambda<\beta$ for roots of $d$ with real $\beta$. In particular, $\sqrt{\beta-\lambda}>0$. Thus if
$\beta<\pi^2/8$ and $\lambda\geq0$, then $\sqrt{2(\beta-\lambda)}\in(0,\pi/2)$ and the values of the trigonometric functions in (\ref{dd}) are positive, i.e., $d\neq0$. The second statement is proved. The last two statements follow immediately from the implicit function theorem applied to the equation  $d=0$ with $\sqrt \lambda$ replaced by $k$.  One needs only to note that $d$ is analytic in $\beta,k$ in a neighborhood of the point $(\beta,k)=(\pi^2/8,0)$ and $d=0, ~\frac{\partial}{\partial k}d=1/(\sqrt2 \pi),~\frac{\partial}{\partial \beta}d=-2/\pi$ at $(\beta,k)=(\pi^2/8,0)$.
\qed
\\

Denote by $L(a)$ the contour of integration in (\ref{lap}) and by $\Gamma(a)$ the contour obtained by splitting of $L(a)$ in two halves (where Im$\lambda\gtrless0$)  and rotating each half around the point $\lambda=a$ by angle $\pi/4$ to the left (in the direction of the half-plane Re$\lambda<0$). Denote by $\Lambda_{a,b},~0\leq b\leq a,$ the closed region in the complex $\lambda-$plane between $L(a)$  and $\Gamma(b)$.

Note that function $d$ grows exponentially when $\lambda$ goes to infinity in the region $\Lambda_a$. Indeed, one can express the trigonometric functions in (\ref{dd}) through exponents with factors $\pm i$ in the exponents, and the growing terms in this exponential representation of (\ref{dd}) can't cancel each other since they have different pre-exponent factors ($\frac{\sqrt{\lambda}}{\sqrt{(\beta-\lambda)}}$ and $1$, respectively).
To be more exact,
\[
|d(\beta, \sqrt{\lambda})|\geq C|\lambda|^{-1}|e^{\sqrt{2\lambda}}|, \quad \lambda\in\Lambda_{a,0}, ~~\lambda\to\infty.
\]
Taking also into account that $d$ is analytic in $k=\sqrt\lambda,~\lambda\in\Lambda_{a,0},$ vanishing only at $k=\sqrt{\lambda_0(\beta)}$, and that the zero is simple, it follows that
\begin{equation} \label{estd}
\frac{1}{d(\beta, \sqrt{\lambda})}= \frac{q(\beta, \sqrt{\lambda})}{\sqrt\lambda-\sqrt{\lambda_0(\beta)}}, \quad |q|\leq(1+|\lambda|)^{3/2}|e^{-\sqrt{2\lambda}}|,  \quad \lambda\in\Lambda_{a,0},
\end{equation}
where $q$ is analytic in both arguments, real when $\lambda$ is real, and $q(\beta,0)>0$.

From (\ref{uu}), (\ref{uu2}), (\ref{estd}), it follows that $u$ can be represented as
\begin{equation} \label{defu}
u= \frac{[h(\beta, \sqrt{\lambda})+h_1(\beta, \sqrt{\lambda},r)]e^{-\sqrt{2\lambda}r}}{(\sqrt\lambda-\sqrt{\lambda_0(\beta)})r}, \quad |h|+|h_1|\leq C(1+|\lambda|)^{3/2},  \quad \lambda\in\Lambda_{a,0},
 \end{equation}
where $h,h_1$ are analytic in $(\beta, \sqrt{\lambda})$, $h_1=0$ when $r>1$, $h$ is real when $\lambda$ is real, and $h(\beta,0)>0$.

Since
function $u$ decays exponentially at infinity in $\Lambda_{a,0}$ and is analytic in $\lambda$ between $L_a$ and $ \Gamma_a$, one can replace contour of integration in (\ref{lap})  by $\Gamma(a)$, i.e.,
\begin{equation} \label{lap1}
p_\beta(t,0,x)=\frac{1}{2\pi i}\int_{\Gamma(a)}e^{t\lambda}u(\lambda,r)d\lambda, \quad r=|x|.
\end{equation}

For any function $w=w(x)$, denote by $w^{(\nu)}$ its moment of order $\nu$, i.e., $w^{(\nu)}=\int_{\mathbb{R}^3}|x|^\nu wdx$, and
\begin{equation} \label{mom}
p_\beta^{(\nu)}=\frac{1}{2\pi i}\int_{\mathbb{R}^3}\int_{\Gamma(a)}r^\nu e^{t\lambda}u(\lambda,r)d\lambda dx, \quad  |\beta-\beta_{\rm cr}|\leq \varepsilon.
\end{equation}
In fact, we will use only $\nu=0$ and $2$.

We will split the region $|\beta -\beta_{\rm cr}|\leq\varepsilon,~t\gg 1, $ into three different subregions and estimate $p_\beta^{(\nu)}$ in each of them separately. First, we assume that
\begin{equation} \label{fc}
\beta_{\rm cr}\leq \beta \leq \beta_{\rm cr}+\varepsilon,~~(\beta - \beta_{\rm cr})\sqrt t\geq 1.
 \end{equation}
We will use the notation
\[
\gamma=\sqrt{\lambda_0(\beta)}.
\]

The integrand in (\ref{mom}) is meromorphic in $\lambda$ between the contours $\Gamma(a)$ and $\Gamma(\frac{1}{16t})$ with the only pole at $\lambda=\lambda_0(\beta)$ (it is located between the contours due to (\ref{lest})). The integrand decays exponentially in $\lambda\in \Lambda_a$ at infinity. Hence the contour $\Gamma(a)$ can be moved to $\Gamma(\frac{1}{16t})$ if we take into account the contribution $v^{(\nu)}_1$ from the pole. Thus $p_\beta^{(\nu)}=v^{(\nu)}_1+v^{(\nu)}_2$, where $v^{(\nu)}_2$ is given by (\ref{mom}) with $\Gamma(a)$ replaced by  $\Gamma(\frac{1}{16t})$.

In order to estimate $v^{(\nu)}_2$, we represent it in the form $v^{(\nu)}_2=w+w_1$ by splitting the terms $h+h_1$ in (\ref{defu}) and writing
the integrals containing these terms separately. We  will use (\ref{defu}) and the existence of a constant $c'>0$ such that
\begin{equation} \label{111}
\frac{1}{|\sqrt\lambda-\gamma|}\leq\frac{c'}{\gamma}, \quad \lambda\in \Gamma(\frac{1}{16t}).
\end{equation}
 The latter inequality follows from (\ref{lest}) and the assumption (\ref{fc}). Since $h_1=0$ for $r>1$, from (\ref{defu}), (\ref{111}) and the substitution $\lambda=\mu/ t$ it follows that
\begin{equation} \label{u1}
|w_1|\leq \frac{C}{\gamma}\int_{\Gamma(\frac{1}{16t})}(1+|\lambda|)^{3/2}|e^{t\lambda}d\lambda |=\frac{C}{t\gamma}\int_{\Gamma(\frac{1}{16})}(1+\frac{|\mu|}{t})^{3/2}|e^{\mu}d\mu |\leq\frac{C_1}{t\gamma}.
\end{equation}

By evaluating the exterior integral in the term $w$ followed by the substitution $\lambda\to \mu/t$, we obtain
\[
w=\frac{1}{2\pi i}\int_{\mathbb{R}^3}\int_{\Gamma(\frac{1}{16t})}\frac{r^{\nu-1}h(\beta, \sqrt{\lambda})e^{\lambda t-\sqrt{2\lambda}r}}{\sqrt\lambda-\gamma}d\lambda dx=\frac{(\nu+1)!}{2\pi i}\int_{\Gamma(\frac{1}{16t})}\frac{h(\beta, \sqrt{\lambda})e^{\lambda t}d\lambda}{(2\lambda)^{\frac{\nu+2}{2}}(\sqrt\lambda-\gamma)}
\]
\[
=\frac{(\nu+1)!t^{\nu/2}}{\gamma2\pi i}\int_{\Gamma(\frac{1}{16})}\frac{h(\beta, \sqrt{\mu/t})e^{\mu}d\mu}{(2\mu)^{\frac{\nu+2}{2}}(\frac{\sqrt\mu}{\gamma\sqrt t}-1)}=c\frac{t^{\nu/2}}{\gamma}(h(\beta,0)+O(\frac{1}{\sqrt t})), \quad t\to \infty,
\]
where
\[
c=c(\gamma\sqrt t)=\frac{(\nu+1)!}{2\pi i}\int_{\Gamma(\frac{1}{16})}\frac{ e^{\mu}d\mu}{(2\mu)^{\frac{\nu+2}{2}}(\frac{\sqrt\mu}{\gamma\sqrt t}-1)}.
\]
Here $\gamma\sqrt t\geq 1/2$ due to (\ref{fc}) and (\ref{lest}). Then one can show that $0 < c_0^- \leq c \leq c_0^+$. From here and (\ref{u1}) it follows that
\begin{equation} \label{v2}
v^{(\nu)}_2=c\frac{(\gamma\sqrt t)^{\nu}}{\gamma^{\nu+1}}(h(\beta,0)+O(\frac{1}{\sqrt t})), \quad t\to \infty.
\end{equation}

Let us evaluate $v^{(\nu)}_1$. Formula (\ref{defu}) implies that the residue $\hat{u}$ of function $u$ at the pole $\lambda=\gamma^2$ is given by
\[
\hat{u}=2\gamma r^{-1}[h(\beta, \gamma)+h_1(\beta, \gamma,r)]e^{-\sqrt{2}\gamma r}.
\]
Since $h_1=0$ when $r>1$ we have
\[
v^{(\nu)}_1=  e^{\gamma^2t}\int_{\mathbb{R}^3}r^\nu\hat{u}dx= 2\gamma e^{\gamma^2t}[ h(\beta,\gamma)\int_{\mathbb{R}^3}r^{\nu-1}e^{-\sqrt{2}\gamma r}dx+O(1)]= e^{\gamma^2t}[\frac{c_\nu h(\beta,\gamma)}{\gamma^{\nu+1}}+O(1)],
\]
where $c_\nu=2\int_{\mathbb{R}^3}r^{\nu-1} e^{-\sqrt {2}r}dx>0$ and the remainder term is uniformly bounded in the region (\ref{fc}). The latter relation with (\ref{v2}) implies that
\[
p_\beta^{(\nu)}\approx\frac{e^{\gamma^2t}}{\gamma^{\nu+1}}, \quad  \beta_{\rm cr}\leq \beta \leq \beta_{\rm cr}+\varepsilon,~~(\beta - \beta_{\rm cr})\sqrt t\geq 1,
\]
and this justifies the validity of the first line in (\ref{th}).

Consider now the case when
\begin{equation} \label{cond2}
|\beta-\beta_{\rm cr}|\leq \varepsilon, \quad |\beta-\beta_{\rm cr}|\sqrt{ t}\leq1.
\end{equation}
 Using (\ref{defu}) we split the integral (\ref{mom}) into two terms, containing $h$ and $h_1$, respectively. Since $h_1=0$ when $r>1$, the term with $h_1$ is uniformly bounded when (\ref{cond2}) holds. Hence,  after the change of the variables $(\lambda,x)\to(\mu/t, z\sqrt t)$ in the first term, we obtain
\begin{equation} \label{intuC}
p_\beta^{(\nu)}=\frac{t^{(1+\nu)/2}}{2\pi i}\int_{\mathbb{R}^3}\int_{\Gamma(a/t)}\frac{|z|^{\nu-1} e^{\mu-\sqrt{2\mu}|z| }}{(\sqrt\mu-\gamma\sqrt t)}h(\beta,\sqrt{\mu/t})d\mu dz+O(1).
\end{equation}
The integrand in (\ref{mom}) is analytic in $\lambda$ in the region on the right from $\Gamma(a)$, and therefore the integrand in (\ref{intuC}) is analytic in $\mu$ to the right of $\Gamma(a/t)$. Since the latter integrand decays exponentially at infinity between $\Gamma(a/t), t\geq 1,$ and $\Gamma(a)$, we can replace the contour of integration in (\ref{intuC}) by $\Gamma(a)$. Thus, if (\ref{cond2}) holds, we can evaluate the exterior integral and rewrite $p_\beta^{(\nu)}$ in the form
\begin{equation} \label{intuD}
p_\beta^{(\nu)}=t^{(1+\nu)/2}[c_1 h(\beta,0)+O(\frac{1}{\sqrt t})], \quad c_1=c_1(\gamma\sqrt t)=\frac{(\nu+1)!}{2\pi i}\int_{\Gamma(a)}\frac{ e^{\mu}d\mu}{(2\mu)^{\frac{\nu+2}{2}}(\sqrt\mu-\gamma\sqrt t)} .
\end{equation}
From (\ref{cond2}) and (\ref{lest}) it follows that $\gamma\sqrt t\leq1$, and therefore one can show (recall also that $a>\gamma^2t$) that there exists constants
$c_0^-$, $c_0^+$ such that $0 < c_0^- \leq c_1 \leq c_0^+$. This justifies (\ref{th}) under assumption (\ref{cond2}).

 The same argument can be applied to prove the theorem in the case  $\beta_{\rm cr}-\varepsilon\leq\beta<\beta_{\rm cr},~(\beta-\beta_{\rm cr})\sqrt t\leq-1$. One needs only to take $|\gamma|\sqrt t$ out of the
integral in (\ref{intuC}) and write (\ref{intuD}) in the form
 \[
p_\beta^{(\nu)}=\frac{t^{(1+\nu)/2}}{|\gamma|\sqrt t}[c_2 h(\beta,0)+O(\frac{1}{\sqrt t})], \quad c_2=c_2(\gamma\sqrt t)=\frac{(\nu+1)!}{2\pi i}\int_{\Gamma(a)}\frac{ e^{\mu}d\mu}{(2\mu)^{\frac{\nu+2}{2}}(\frac{\sqrt\mu}{|\gamma|\sqrt t}+1)} .
\]
Then $|\gamma|\sqrt t\geq \frac{1}{2}$ and there there exists constants
$c_0^-$, $c_0^+$ such that $0 < c_0^- \leq c_2 \leq c_0^+$. This justifies (\ref{th}) in the latter case.
\qed
\\
\\\\
\noindent {\bf \large Acknowledgments}:
The work of  L. Koralov was supported by the ARO grant W911NF1710419. The work of S. Molchanov was supported by the NSF grant DMS-1714402 and by the Russian Science
 Foundation,  project ${\rm N}^o$ 17-11-01098 and project ${\rm N}^o$~20-11-20119.  The work of B. Vainberg was supported by the NSF grant DMS-1714402 and the Simons
 Foundation grant 527180.
\\
\\

\end{document}